# Title: Near-field optical random mapping (NORM) microscopy.

**Authors:** Yuri V.Miklyaev[1]*, Sergey A. Asselborn[2], Konstantin A. Zaytsev[2], Maxim Ya. Darscht[1]

**Affiliations:**

[1]South Ural State University, 454080, Lenin Ave., 76, Chelyabinsk, Russia.

[2]Institute of Electrophysics, Ekaterinburg, Russia.

*Correspondence to: mikliaevyv@susu.ac.ru

**Abstract**: In recent years several methods to overcome diffraction limit in the far field microscopy have been demonstrated. Still the problem of superresolution is reliably solved only for fluorescent microscopy, giving a resolution of up to 20-30nm. Obtaining the optical resolution lower than 100nm without fluorescent dyes requires using rather slow and complicated technique of scanning near filed optical microscope (SNOM). We propose and demonstrate a method of optical near field acquisition by far-field microscope through observation of nanoparticles Brownian motion in immersion liquid. The resolution of the method is restricted only by the size of nanoparticles that can be registered (detected) by a given far field microscope. From this point of view this resolution can achieve up to 10-20nm. Up to now we achieved a resolution of about 140nm observing 120nm particles through an objective with N.A.=0.4. The resolution is thus improved by factor of five for a given microscope objective.

**One Sentence Summary:** We propose and demonstrate a method of optical near field acquisition by far-field microscope through observation of nanoparticles Brownian motion in immersion liquid.

**Main Text:**

Resolution is one of the most important parameter in microscopy. In spite of much higher resolution of electron and scanning (atomic and tunneling) microscopy these methods do not give information about optical field distribution on the object surface. An exception from this rule is scanning near field optical microscopy (SNOM) (*1,2*). With SNOM one can acquire light intensity distribution and radiation spectra with resolution much higher than Rayleigh criterion allows. But scanning approach and necessity to keep a sensor tip not farther than 50-100nm from an object surface result in low speed and higher price of such a devices.

Although many methods of superresolution in far field microscopy were developed in past decade, most revolutionary of them are based on fluorescence microscopy (*3-9*). In such a way one only gets the information about the dye distribution on\in object, but such information is very important for biological and medical applications. These methods can be loosely divided on deterministic (STED, SIM, SSIM-microscopy) and stochastic (PALM, STORM etc.) approaches.

Method of stimulated emission depletion (STED) (*3,4,10,11*) is based on sample scanning by focused laser beam and requires a special dye that can be excited by radiation with one wavelength and depleted by another wavelength. As a result of combination of such two wavelengths, the region of excitation can be reduced in diameter to about 1/30 of wavelength.

Some other methods (*5-8*) rely on such a fluorophores excitation when at each moment luminescence takes place for individual molecules those images are spatially separated. This allows localizing these molecules with high precision.

Quite different philosophy is utilized in technique called by structured illumination microscopy (SIM) (*9*). This method in some sense is analogous to confocal microscopy (*12-14*). But instead beam waist scanning of confocal microscopy here a sequence of sinusoidal patterns is used for object illumination. Like in confocal microscope we get in this case a 3D imaging. With this SIM method one can get resolution improvement by factor of 2 (for comparison, in confocal microscope a resolution improvement is equal to factor of SQR(2)). Such a resolution enhancement is allowed not only for fluorescent samples, but also for phase and scattering objects (*15-17*).

SIM of non-fluorescent objects is limited to this factor-of-two improvement. Utilizing non-linear features in excitation and emission of fluorescence in combination with SIM can give much higher resolution. This technique is called as saturated SIM (SSIM) (*18*). The resolution of SSIM scales with the level of saturation. For example, 50nm lateral resolution was reported using fluorescent beads (*18*).

Recently a new phenomenon called superoscillation was investigated in relation to subdiffraction resolution (*19-22*). The same origin of superresolution has a technique based on eigenmode decomposition (*23*). The main shortcoming and distinguishing feature of this concept is that an intensity of resolved details exponentially decays with increasing of observed area (field of view).

To convert a near-field image into the far-field, a variety of so called far-field superlenses (FSL) and hyperlenses were proposed and demonstrated. Some of them are based on the resonant coupling of evanescence waves into surface plasmon polaritons (*24-28*). Another class of FSL is based on subwavelength imaging and magnification by dielectric lenses or spheres with curvature radius of several micrometers. Small scale intensity distributions can be achieved with very small foci. Being incorporated into an optical microscope this approach provides a resolution of up to 50nm (*29-31*).

In this work we demonstrate a new method to recover an image with subwavelength resolution. The main principle of this method was proposed by us in (*32*) and can be explained as follows. To acquire super-high spatial frequencies of an image we place the object into a scattering medium. The medium scatters (diffracts) high spatial harmonics enclosed in evanescent waves into propagating waves. It is possible to reconstruct the information about near-field from far-field image if we know a spatial harmonics content of the scattering medium. It is obvious, that such a scattering medium must contain a subwavelength spatial harmonics. Apparent example of such a medium is a suspension of nanoparticles. It can be transparent, reflecting or absorbing nanoparticles of any form but must be subwavelength in size. In this case to recover the near-field image from far-field it is sufficient to determine scatterer distribution.

We utilize object surface scanning by stochastic moving nanoparticles. The process of image acquisition can be described as follows. Sample surface is observed by ordinary optical microscope. On the sample we place a suspension of nanoparticles with the size much lower than the optical resolution of the microscope. Through the microscope nanopaticles look like symmetric round spots. Utilizing image processing we can measure lateral coordinates of the given particle with a precision much higher than the resolution of the microscope. In this case the detected brightness of the corresponding spot is defined by the local near field in the point where the particle is located. From a single frame (image) we can get information about the local intensity of the near field in several points chaotically distributed on the sample surface. Due to the Brownian motion in the subsequent (in the next frame) we get another distribution of particles over the sample. Collecting the information from many frames we can map out the near field intensity distribution across the whole field of view of the microscope (we can stochastically scan all sample surface that we see in the microscope).

It should be pointed out that such constructed near field map will be drawn by particles moving at different altitudes above the sample surface (but inside of the depth of focus). As it is well known from the ordinary (traditional) SNOM technique, the distance between a surface and a sensor defines the minimum feature size that can be resolved. In other words, farther floating particles can draw only low spatial frequencies of the image. Due to this fact the resulted image will contain disproportionately strong low frequencies. As a result we will get image blurring. To exclude the influence of such "waste" particles it is sufficient to make a Fourier filtering by damping low frequencies.

The proposed method of the near field acquisition can be realized in a reflecting or a transmission mode.

For the experimental verification of the proposed technique we used the conventional microscope LOMO BioLAM I, operating in a reflection mode. A water suspension of TiO2 nanoparticles was placed between object surface and a cover glass. Nanoparticles were observed through a 34X objective lens with a numerical aperture (NA) of 0.4 with LED illumination at 550nm wavelength and a corresponding point spread function (PSF) size of about 840nm.

To get a larger contrast of the light scattered by particles we used dark field mode of the illumination. Average size of nanoparticles was about 120nm. As a test samples we have taken Cr grating on a glass consisting of three 400nm-wide lines, spaced 1.6μm apart and 2D array of 100nm holes with 400nm pitch in a Ti film on a glass. To avoid adhesion of particles to the object surface we added a surfactant in a suspension. The picture of moving particles was taken by CCD camera with sensitivity 1.4 nW/mm² at 120Hz.

The resulted movie was analyzed by our own tracking software. Basically our method is analogous to the algorithm proposed in (*33*). But to get a higher processing speed with a sufficient precision we developed and combined some different approaches. Image filtering was made by sum of absolute difference (SAD) with Gaussian (*34*). SAD is an extremely fast metric due to its simplicity and easily parallelizable. It is implemented in processor unit utilizing such instructions as MMX and SSE2. Local maxima search algorithm was taken from (*35*). To locate a particle we used direct fits of Gaussian curves to the intensity profile of an isolated particle. As it was shown (*36*), for the case of a sub-wavelength particle, such an approach is the best one in terms of an accuracy and a precision, and is the most robust at low signal-to-noise ratio. As a result we can make real-time video processing of up to 500 frames/sec (with single-core CPU) for 320X240 pixels per frame. In actual experiment we used an ordinary 8-bit CCD camera with 100 frames/sec acquisition rate. The precision of particle location is about $1/20^{th}$ of the pixel when PSF size is equal to 7 pixels and signal-to-noise ratio is near 10.

Calculated particle coordinates were rounded to 1/3 pixel, i.e. each pixel of initial image was divided on to 3X3 subpixels. The resulting intensity at a given subpixel was calculated as a sum intensity of all particles over all frames with coordinates located at this subpixel. In our case a CCD pixel size was 4.65μm. Taking into account magnification factor of 34 we find that one subpixel corresponds to 46nm on the object. Instead of using a special microscope chamber to minimize a thermal and a mechanical drift (*37*) we trace the drift by observing a low resolution image shift in time and making a corresponding correction of particle coordinates.

Optical images of Cr triple stripes on a glass are shown on the Figure 1, a to c. It was mentioned above, that not all of the detected particles are in the close proximity to the object surface. It reduces the contribution of high spatial frequencies in the resulted image. To reconstruct the real proportion of different Fourier components we make Fourier filtering of the image (Fig. 1 d). From Figure 1 it can be seen, that NORM made through 0.4NA gives higher definition than conventional imaging with 1.0NA.

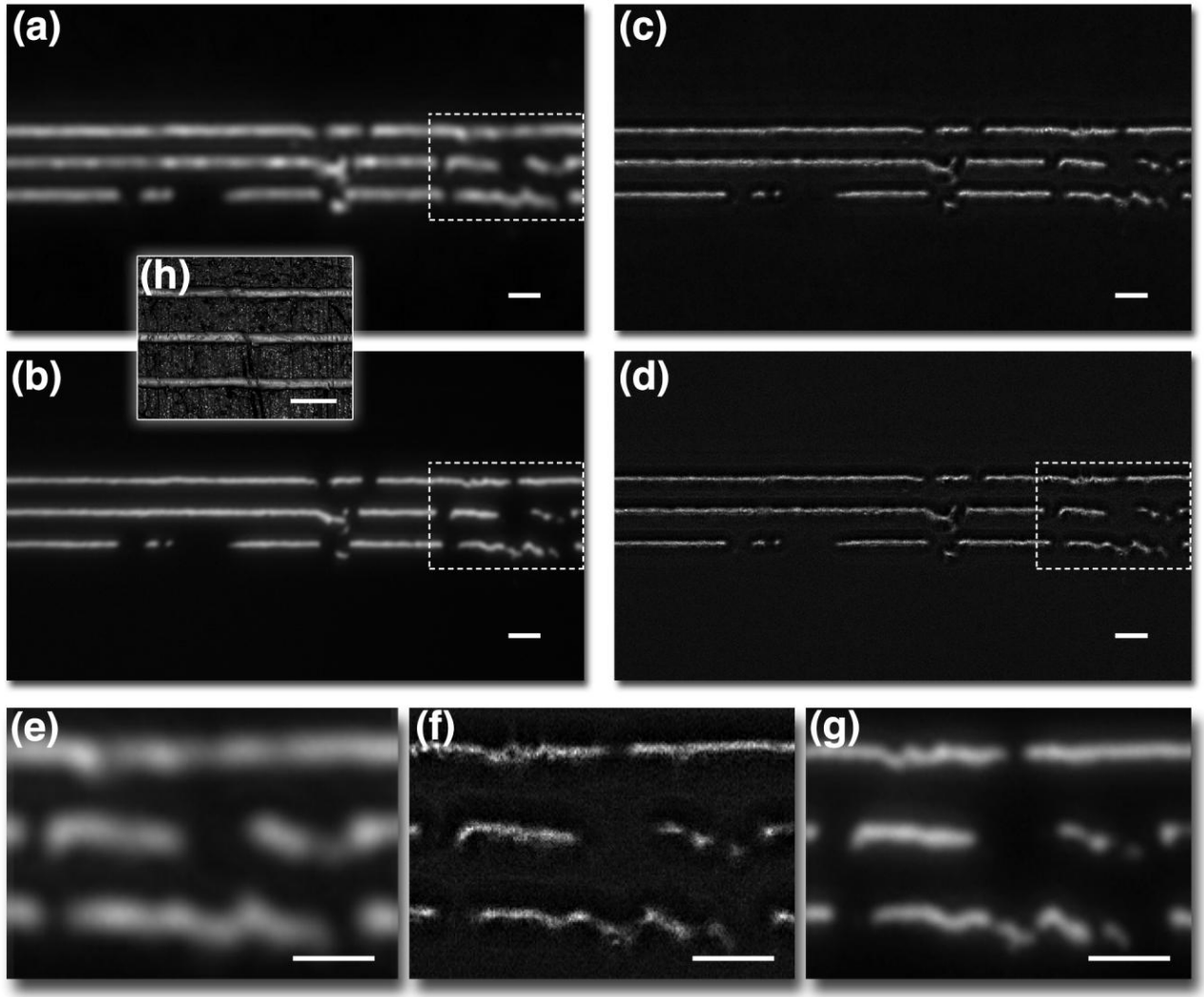

**Fig.1.** Optical images of Cr stripes with different acquisition methods. (**a**) objective with 0.4NA, dark field illumination mode; (**b**) immersion objective 1.0NA, light field illumination mode; (**c**) NORM without Fourier filtering through 0.4NA; (**d**) NORM after Fourier filtering, (**e**) magnified part of image from (a), (**f**) magnified part of image from (d), (**g**) magnified part of image from (b), (**h**) AFM image of Cr stripes on glass substrate. Scale bars, 2 μm. Image acquisition time by NORM was about 5 min, but all details were drawn already after 1.5 min.

The Fourier transforms of all the three images are shown on the Figures 2, a to c. From these graphs we can conclude that the resolution of NORM improves the resolution of the objective in use by the factor of 5 and exceeds the resolution of the objective with 1.0NA by the factor of 2.

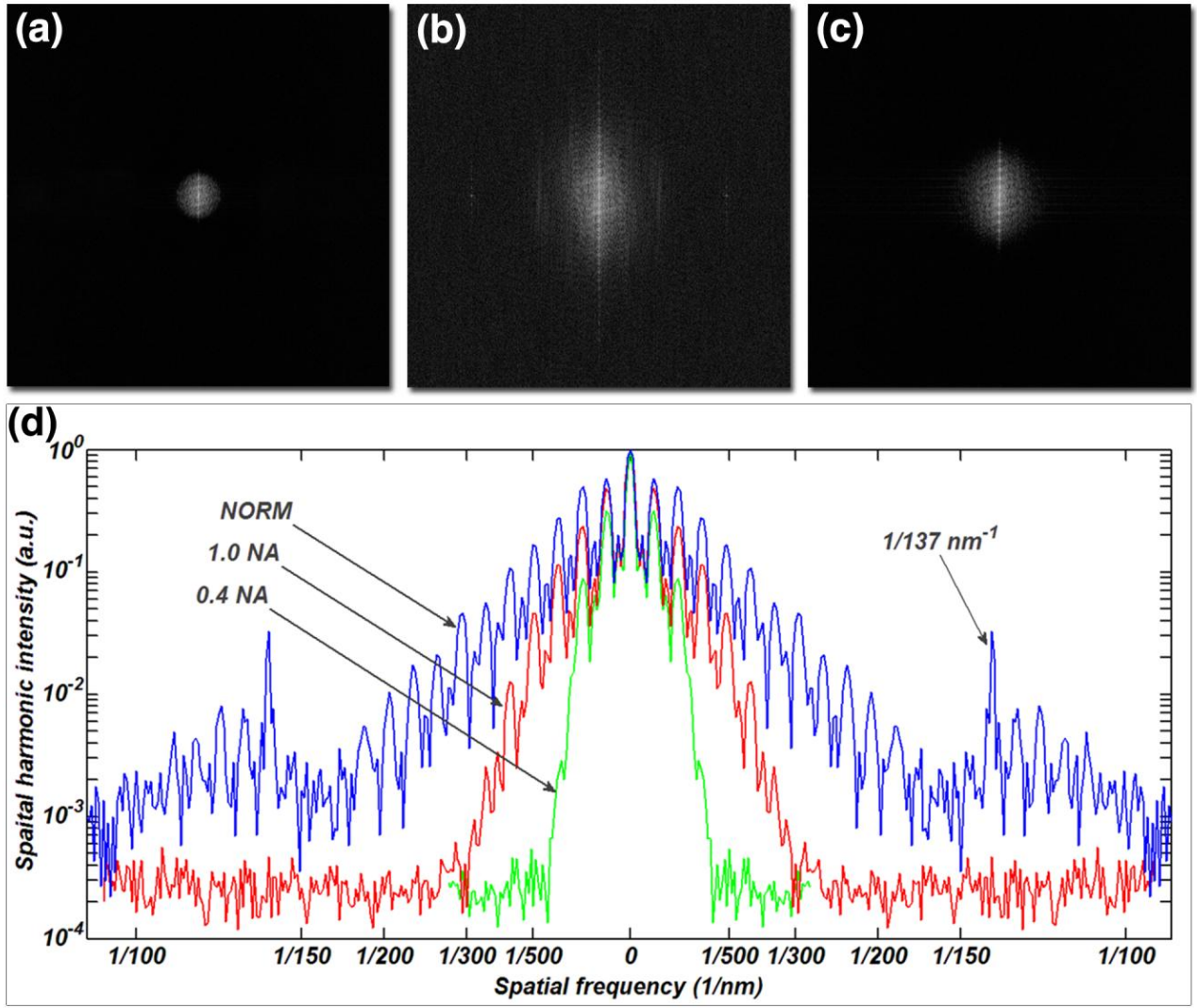

**Fig.2.** The Fourier transforms of the three optical images obtained by three different methods: (**a**) and green curve is the objective with 0.4NA in the dark field illumination mode; (**b**) and blue curve is the NORM (without Fourier filtering); (**c**) and red curve is the immersion objective with 1.0NA. (**d**) Fourier transform cross-section of the three images along the y-axis. In the Fourier transform of the NORM image there are some narrow peaks at frequencies of $1/137$ nm$^{-1}$. These peaks are particle tracking algorithm artifacts and correspond to periods equal to the single pixel of initial image. For objectives with 0.4NA and 1.0NA the maximum spatial frequencies have periods 630 nm and 280 nm, correspondingly. It is near from the theoretical limit of objectives with such NA's. At the same time the spatial harmonics exceeding the noise level in NORM image have periods of up to 120 nm.

Similar results were obtained for the titanium mask of 100nm holes array (Fig. 3,4) by scanning time equal to 7 min. In this case we do not observe any structure with the 0.4NA objective (Fig. 3, a). It comes from a fact, that the resolution of the objective (630nm) is lower than the raster pitch (400nm). In the immersion objective (1.0NA) we get only the first diffraction order with the period of 400nm. Corresponding image has sinusoidal intensity distribution in X-, Y- coordinates. Second diffraction order (200nm) has higher frequency than cut-off frequency of the objective (Fig.

4, c, red curve). For NORM picture we get second and third diffraction orders (Fig. 4, c, blue curve) with periods 200nm and 133nm.

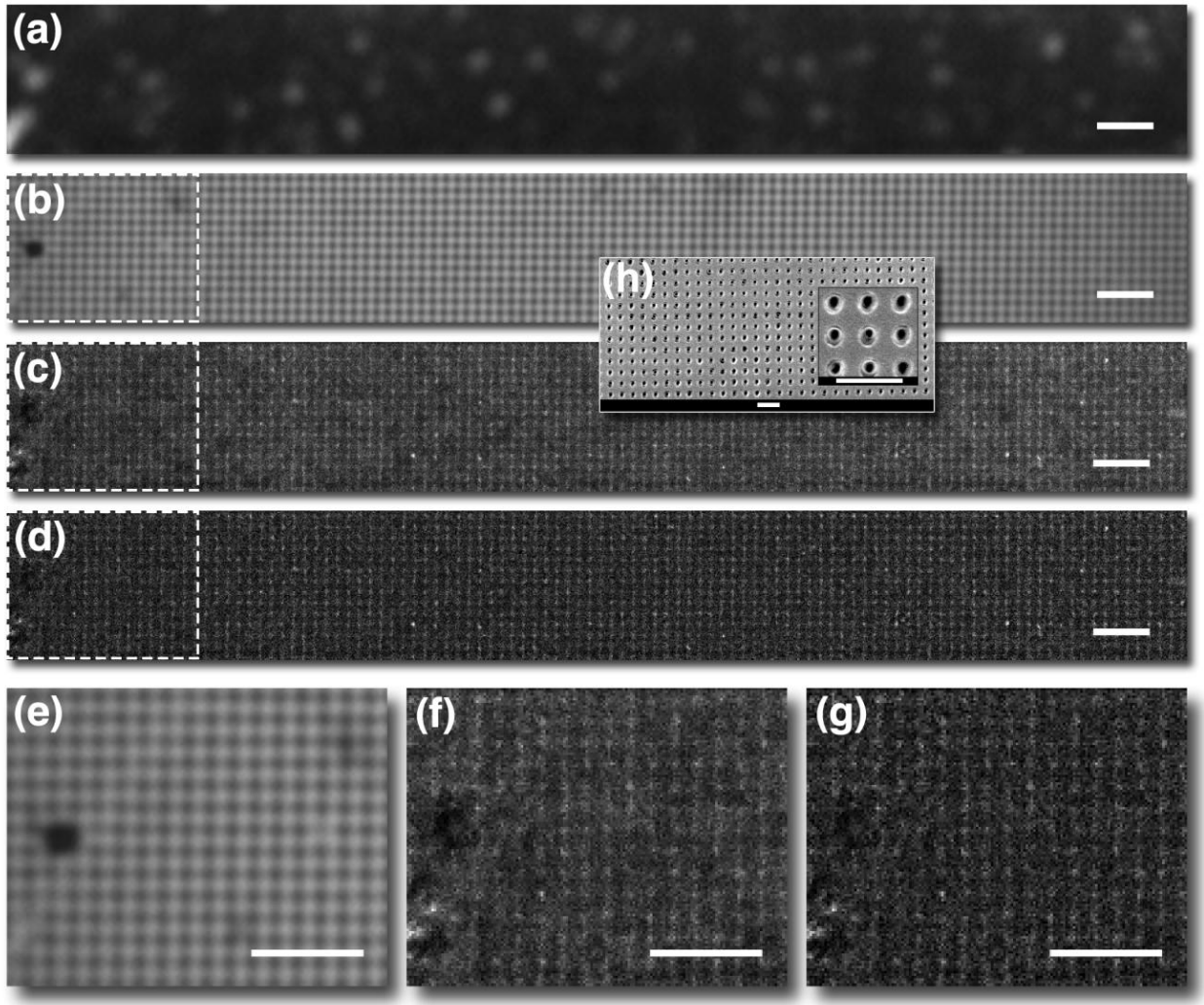

**Fig.3.** Optical images of Ti mask with holes by different acquisition methods. (**a**) objective with 0.4NA, dark field illumination mode. Bright spots here are images of nanoparticles with PSF equal to 840nm; (**b**) immersion objective 1.0NA, light field illumination mode; (**c**) NORM without Fourier filtering through 0.4NA; (**d**) NORM after Fourier filtering, (**e**) magnified part of image from (b), (**f**) magnified part of image from (c), (**g**) magnified part of image from (d); Scale bars, 2μm; (**h**) SEM image of Ti mask with holes on glass substrate. Scale bar 800 nm. It is worth to mention one distinctive feature of the image obtained by NORM. Holes look like bright spots in contrast to expected dark ones. This fact we interpret by near field subwavelength light focusing due to reflection on conical surface of holes. Such an effect is not visible in the image obtained by 1.0NA objective in the bright field illumination mode, where we see darks spots corresponding to holes.

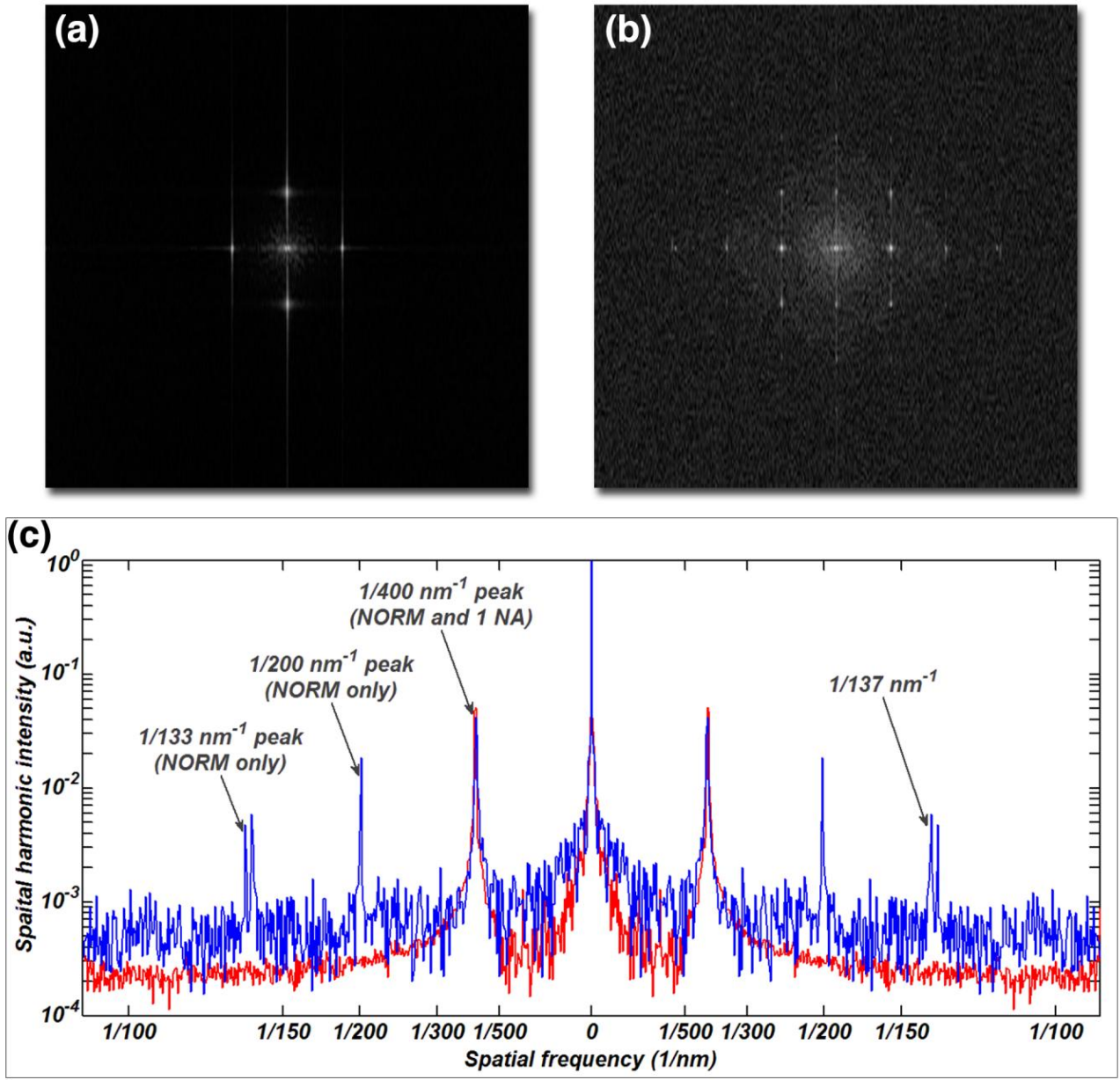

**Fig.4.** Fourier transforms of two optical images obtained by different methods: (**a**) and red curve corresponds to immersion objective with 1.0NA in bright field illumination mode; (**b**) and blue curve corresponds to NORM (without Fourier filtering). (**c**) Fourier transform cross-section of two images along y-axis.

It is worth to mention one distinctive (interesting) feature of the image obtained by NORM with Ti-mask. Holes look like bright spots in contrast to expected dark ones. This fact we interpret by near field subwavelength light focusing due to reflection on conical surface of holes. Such an effect is not visible in the image obtained by 1.0NA objective in the bright field illumination mode, where we see darks spots corresponding to holes.

For further improvement of resolution and image acquisition speed it is worth to use electron multiplying CCD (EMCCD) camera. Due to much higher sensitivity with proper illumination we can observe nanoparticles down to 10nm in size, analogously to a systems for particle size analysis (*38,39*). Also, it is possible to accelerate the video acquisition to 500 frames/sec or higher. With a larger NA objectives and correspondingly lower depth of field and PSF diameter much higher nanoparticle concentrations can be used. According to our estimates it is possible to obtain 20nm resolution images in 1min scanning time.

It is interesting to use the original approach proposed in (*8*), based on the astigmatic imaging of nanoparticles. In this case one can control the altitude of floating particles by measuring the apparent nanoparticle ellipticity. This will allows us to discriminate nanoparticles that are far from the surface and to avoid the image blurring by these particles. Besides, such an analysis of the particle altitude distribution can give information about the surface profile.

Combining the proposed method with TIRF microscopy (*40,41*) allows us to avoid light scattering by particles that are remote from the surface. In this case we will enhance image contrast and will be able to increase the particle concentration together with the acquisition speed.

Utilization of nanobubles instead of nanoparticles can be interesting in those cases where one wants to avoid an object contamination by nanoparticles (*42-44*).

In conclusion, we have demonstrated a new method of superresolution microscopy. Resolution improvement by the factor of five was obtained with a conventional optical microscope and a CCD camera. This method has no restriction in obtaining color images with large depth of focus. We are confident, that the further development of this method can provide resolution up to 20nm with much higher scanning speed. Moreover, combining with other methods of optical microscopy and utilizing different nanoparticles (metallic, magnetic, luminescent, etc.) for scanning extends possibilities for various applications.

**Contributions and acknowledgments:** The principle of super-oscillatory imaging was suggested by Yu.V.M. with contribution from S.A.A.; S.A.A. developed mathematical algorithms for particle tracking and created the particle tracking software. Optical experiment was undertaken by S.A.A. with contribution from K.A.Z.. K.A.Z. implemented the imaging software, performed data post-processing. Yu.V.M. supervised the project and wrote the manuscript with contributions from all co-authors. M.Ya.D. made resolution assessments and comparison with other microscopy methods and nanosuspension preparations. All authors contributed to the discussions of results and planning of experiments.
Authors are grateful to V.S.Pavelyev for mask fabrication and to A.Goun for useful discussions.